\documentclass[a4paper,11pt]{article}
\pdfoutput=1 
\usepackage{jheppub} 
\usepackage[T1]{fontenc} 
\usepackage{graphicx}
\usepackage{placeins}
\usepackage{xspace}
\usepackage{soul}
\usepackage{siunitx}
\usepackage{upgreek}
\usepackage{subcaption}
\usepackage[dvipsnames]{xcolor}

\def\be{\begin{equation}}
\def\ee{\end{equation}}
\def\bea{\begin{eqnarray}}
\def\eea{\end{eqnarray}}

\newcommand{\mceq}{\texttt{MCEq}\,}

\preprint{\begin{flushright} IFIC/23-02 \\ 
\end{flushright}}

\title{\boldmath Testing Heavy Neutral Leptons in Cosmic Ray Beam Dump experiments }
\author[a]{Oliver Fischer,} 
\author[b]{Baibhab Pattnaik,}
\author[b]{Jos\'e Zurita}

\affiliation[a]{Department of Mathematical Sciences, University of Liverpool, \\ Liverpool L69 3BX, United Kingdom}
\affiliation[b]{Instituto de F\'{\i}sica Corpuscular, CSIC-Universitat de Val\`encia, \\ Catedr\'{a}tico Jos\'{e} Beltr\'{a}n 2, E-46980, Valencia, Spain}

\emailAdd{Oliver.Fischer@liverpool.ac.uk}
\emailAdd{Baibhab.Pattnaik@ific.uv.es}
\emailAdd{Jose.Zurita@ific.uv.es}

\abstract{In this work, we discuss the possibility to test Heavy Neutral Leptons (HNLs) using ``Cosmic Ray Beam Dump'' experiments. In analogy with terrestrial beam dump experiments, where a beam first hits a target and is then absorbed by a shield, we consider high-energy incident cosmic rays impinging on the Earth's atmosphere and then the Earth's surface. 
We focus here on HNL production from atmospherically produced kaon, pion and $D$-meson decays, and discuss the possible explanation of the appearing Cherenkov showers observed by the SHALON Cherenkov telescope and the ultra-high energy events detected by the neutrino experiment ANITA. 
We show that these observations can not be explained with a long-lived HNL, as the relevant parameter space is excluded by existing constraints. 
Then we propose two new experimental setups that are inspired by 
these experiments, namely a Cherenkov telescope pointing at the horizon and shielded by the mountain cliff at Mount Thor, and a geostationary satellite 
that observes part of the Sahara desert.
We show that the Cherenkov telescope at Mount Thor can probe currently untested HNL parameter space for masses below the kaon mass. We also show that the geostationary satellite experiment can significantly increase the HNL parameter space coverage in the whole mass range from 10 MeV up to 2 GeV and test neutrino mixing $|U_{\alpha4}|^2$ down to $10^{-11}$ for masses around 300 MeV.
}
\begin{document} 
\maketitle
\flushbottom

\section{Introduction}
\label{s.intro}
The Standard Model of particle physics (SM) is a highly successful theory over many energy scales. Yet, a large number of shortcomings of the SM provide strong support for the existence of physics Beyond Standard Model (BSM): neutrino masses, dark matter, the strong CP problem, just to name a few. 
The Large Hadron Collider (LHC) has carried out an extensive campaign to probe a large palette of new physics scenarios at the electroweak and TeV scale. 
One of the possible options that received a lot of attention in recent years is the search for so-called Long-Lived Particles (LLPs). Such particles appear naturally in many BSM scenarios that address subsets of the open questions in particle physics~\cite{Curtin:2018mvb}.
A prominent laboratory for LLP searches is the LHC, which can test a large suite of LLP-related signatures, cf.\ e.g.\ ref.~\cite{Lee:2018pag} for a review. It is worthwhile to note that LHC searches often require LLPs to have masses of at least ${\cal O}(10)$~GeV, such that they become distinguishable from hadronic noise.
LLPs with masses in the sub-GeV region are generally difficult to test in the high-energy environment of the LHC and therefore might escape detection, in particular when their lifetimes exceed the detector dimensions.
LLPs that are light and have very long lifetimes are targeted by the new external LHC detectors FASER~\cite{Feng:2017uoz}, SND@LHC \cite{Boyarsky:2021moj} and CODEX-b \cite{Aielli:2019ivi}, and furthermore by the proposed detectors MATHUSLA \cite{Curtin:2018mvb}, ANUBIS \cite{Bauer:2019vqk}, AL3X \cite{Gligorov:2018vkc}, and others \cite{Alimena:2019zri}. Moreover, LLP searches beyond the LHC are also possible in other dedicated experiments, see e.g.~\cite{Beacham:2019nyx}.
Another natural laboratory to search for LLP signatures is neutrino experiments, where mesons (e.g.\ neutral and charged $\pi$, K, D) are copiously produced from colliding a proton beam with a target, 
which then decay into muons and neutrinos. 
Subsequently, this neutrino beam is observed
downstream with dedicated detectors.

Similarly to producing neutrinos, meson decays can produce LLPs with sub-GeV masses via rare decays and thus create an LLP beam. This beam could in principle also be tested with downstream detectors, for instance via LLP decays into visible final states, see e.g.\ this MiniBooNE dark matter search in a proton beam dump \cite{Aguilar-Arevalo:2019wki}.

A third possibility exists to test LLPs with masses of at most ${\cal O}(1)$~GeV: cosmic rays impinging on the atmosphere give rise to showers of SM particles that can decay before reaching the ground and thus create a continuous background of LLPs with energies ranging from few GeV up to EeV.
Such atmospherically produced LLPs could be tested with large-volume neutrino detectors placed below the Earth surface, such as Super Kamiokande or IceCube, see e.g.~\cite{Kusenko:2004qc,Asaka:2012hc,Masip:2014xna,Arguelles:2019ziu,Coloma:2019htx,Meighen-Berger:2020eun,Su:2020zny,ArguellesDelgado:2021lek,Gu:2021lni,Candia:2021bsl,Undagoitia:2021tza,Iguro:2021xsu,Arguelles:2022fqq,Gustafson:2022rsz,Darme:2022bew,Cui:2022owf,Cheung:2022umw}.

In this work, we zoom in on the fact that, for cosmic rays impinging on the Earth, the atmosphere and the Earth surface act as a beam dump for the incident cosmic ray beam, also copiously producing heavy mesons which can act as a source of Long-Lived Particles, which further decay into highly energetic SM particles. 
We will focus here on two experiments with unexpected excesses in their datasets, and analyze if those anomalies can be explained in terms of Long-Lived Particles. On one side, we will study the Cherenkov telescope SHALON, which observed five showers at TeV energies that were originally interpreted as the decays of a heavy neutral lepton (HNL) \cite{Sinitsyna:2009dn}, although it was concluded that the required production rate seems to be too high. On the other side, we consider the balloon-borne neutrino experiment ANITA, which observed two Earth-emergent ultrahigh-energy events that cannot be explained within the SM, see e.g.~\cite{Fox:2018syq,Collins:2018jpg,Hooper:2019ytr,Cline:2019snp,Heurtier:2019rkz,Borah:2019ciw,Anchordoqui:2019utb,Abdullah:2019ofw,Esmaili:2019pcy,Altmannshofer:2020axr,BhupalDev:2020zcy,Liang:2021rnv}.

For concreteness, we consider the well-studied case of Heavy Neutral Leptons (HNLs, see e.g.~\cite{Abdullahi:2022jlv} for a recent review), which is appealing as it provides a minimal theoretical setup that can accommodate neutrino data (from oscillations and the atmosphere), explaining the observed mass and mixing patterns. Nonetheless, we stress that our considerations are fairly general and can also be applied to other sub-GeV neutral LLP scenarios, such as dark photons or axion-like particles.

The current work is structured as follows. In section~\ref{s.theory} we review the basic phenomenological features of the HNL model, along with the calculation for the production, decay and detection of atmospherically produced long-lived HNLs. We also sketch the detection strategy for an earth-based Cherenkov telescope and for a satellite experiment, and comment on possible background sources. In section~\ref{s.sens} we interpret the current SHALON and ANITA dataset in the context of HNLs, disproving the possibility that those excesses are due to long-lived HNLs. These sensitivity calculations lead us to speculate on the possible modifications to these baseline designs, and we then compute the sensitivity of two new experiments: a Cherenkov telescope located in Mount Thor in Canada (`MtThor') and a particle detector in geostationary orbit examining a ``zero-background'' environment such as the Sahara desert, 
having HNLs as the fundamental physics case.  We reserve section~\ref{s.conclu} for our conclusions.

\section{Theory framework}
\label{s.theory}
Here we focus on the class of models that explain the observed neutrino oscillations by extending the SM via the type-I seesaw mechanism with right-handed neutrinos \cite{Minkowski:1977sc,Mohapatra:1979ia,Schechter:1980gr}.
In these models, typically $n\geq 2$ right-handed neutrinos $N_i$ are introduced that have Yukawa interactions with the lepton doublets and the Higgs doublet.
After breaking of the electroweak symmetry this leads to neutrino mixing, which gives rise to the so-called na\"ive type-I seesaw relation for the active-sterile mixing angle
\begin{equation}
    \theta = \sqrt{\frac{m_\nu}{M}}\,,
    \label{eq:seesaw}
\end{equation}
where $m_\nu$ and $M$ are the masses of the light and heavy neutrino mass eigenstates, respectively. An overview of this class of models and their phenomenology was presented in ref.~\cite{Drewes:2013gca}.

In general, the model parameters can be reconstructed via a ``bottom-up approach'' from the observed pattern of masses and mixings in the neutrino sector \cite{Casas:2001sr}.
Such models frequently predict lepton-number-violating signatures such as neutrinoless double beta decay \cite{Schechter:1981bd} or the decays $\mu \to e\gamma$ \cite{Minkowski:1977sc} that have not been observed to date, thus strictly limiting the parameter space.

\subsection{Toy model}
Realistic neutrino models reproduce the measured mass spectrum and mixings by fitting a large number of parameters to the available experimental data.
Certain scenarios exist, however, wherein additional symmetries alleviate some of the constraints, such that in particular the neutrino mixing parameter can be larger than the simple estimate in eq.~\eqref{eq:seesaw} \cite{Kersten:2007vk,Gavela:2009cd}.

We here consider a model in a symmetry-protected scenario and parameterise it with an extension of the SM with a single right-handed neutrino $N_R$.
This allows us to add the following terms to the Lagrangian:
\begin{equation}
{\cal L} = {\cal L}_{SM} + i \overline{N_R} \gamma^\mu \partial_\mu N_R - Y_{\alpha} \overline L_\alpha \tilde \Phi N_R - M \overline{N^c_R} N_R + \text{ h. c.}
\end{equation}
Here ${\cal L}_{SM}$ includes the SM field content; $Y$ is a Yukawa vector that quantifies the Yukawa couplings between $N_R$, the lepton doublet $L_\alpha$ and the Higgs doublet $\tilde \Phi = i\sigma_2^* \Phi$; and $M$ is the Majorana mass for the right-handed neutrino fields.\footnote{
In the model with exactly one $N_R$ the Majorana mass allows for processes that violate lepton number, such as (a single) light neutrino mass and neutrino-less double beta decay, the latter being one of the most sensitive probes of the Majorana nature of the light active neutrinos, cf.~\cite{Agostini:2022zub} for a recent review. Such effects can be suppressed in models with more than one HNL that are quasi-Dirac states with an associated small mass splitting~\cite{Kersten:2007vk}. In the following, we assume that such a suppression is given in our model and we will not discuss lepton number violation further.}
The above Lagrangian leads to the type-I seesaw mechanism for neutrino mass generation, which has been widely discussed in the literature (see e.g.\ \cite{Minkowski:1977sc}).

The mixing of the $3+1$ neutral fermions can be parameterised by the extended PMNS  matrix $U$ and leads to two massless and one light neutrino mass eigenstates $\nu$ (which are mostly left-chiral) and one heavy neutrino mass eigenstate $N$ (which is mostly right-chiral).
Here we focus on the mixing between left-handed and right-handed neutrinos often expressed as $U_{\alpha j}$, which mixes neutrino with flavor $\alpha$ with the right-handed neutrino $N_{R_j}$, and as we only have one right-handed neutrino in our model it is $j=4$.
This mixing allows the heavy mass eigenstates to partake in the weak interactions $W^{\mp\, \mu} j^\mp_\mu,\, Z_0^\mu j_\mu^0$ via the currents:
\begin{equation}
    j_\mu^\pm  \supset   \frac{g}{2} \, U_{\alpha 4} \, \bar \ell_\alpha \, \gamma_\mu P_L N   \,, \qquad 
j_\mu^0 = \frac{g}{2\,c_W} \sum\limits_{i,j=1}^{4} \vartheta_{ij} \overline{ \tilde n_i} \gamma_\mu P_L \tilde n_j\,.
\end{equation}
Here, $g$ and $c_W$ are the $SU(2)_L$ coupling constant and the cosine of the Weinberg angle respectively, $\ell_\alpha$ is a lepton of flavor $\alpha$, $P_L$ is a projection operator, and summation over repeated indices is understood. Moreover we followed \cite{Antusch:2015mia} and introduced 
\begin{equation}
\vartheta_{ij} = \sum_{\alpha=e,\mu,\tau} U^\dagger_{i\alpha}U_{\alpha j}^{}\,,
\end{equation}
with $U$ being the extended PMNS mixing matrix as mentioned above, and we introduced
\begin{equation}
    \tilde n_j = \left(\nu_1,\nu_2,\nu_3,N \right)^T_j = U_{j \alpha}^{\dagger} n_\alpha\,.
\end{equation}
Since $N$ interacts weakly and is neutral and heavy, it is often referred to as a `heavy neutral lepton' (HNL). In the following we will use both, `HNL' and `$N$' to denote the heavy neutrino mass eigenstate.
Its interactions are governed by the three mixing parameters $U_{\alpha} \equiv U_{\alpha 4}$ with $\alpha = e, \mu, \tau$. The present experimental status of HNLs and the future prospects were recently summarized in reference~\cite{Abdullahi:2022jlv}.

We remark that while our toy model is somewhat disconnected from the original motivation to explain neutrino oscillations, it remains sufficiently complex to capture the features of HNL from more complicated models. 
The results for this model as shown in section~\ref{s.sens} can be reinterpreted in the more general class of ``realistic neutrino oscillation models'' as discussed e.g.\ in ref.~\cite{Tastet:2021vwp}.

\subsection{Production and decay}
In the present work, we consider HNLs produced from meson decays in atmospheric showers, such that their masses are limited from above by the heavy meson masses.
Hence we focus on HNLs at and below the GeV scale, for which a complete overview of production rates and decay probabilities can be found in~\cite{Bondarenko:2018ptm}. Below we only summarise the features relevant to this work.

The weak interactions allow the production of $N$ from decays of charged mesons $m^\pm$ via the process $m^\pm \to \ell_\alpha^\pm N$ with branching ratio
\begin{equation}
    \text{Br}(m^\pm \to \ell_\alpha^\pm N ) =     \text{Br}(m^\pm \to \ell_\alpha^\pm \nu)|U_{\alpha}|^2\, 
    \frac{\left(x_N^2 + x_\alpha^2-(x_N^2-x_\alpha^2)^2 \right)}{x_\alpha^2(1 - x_\alpha^2)^2} \sqrt{\lambda(1, x_\alpha, x_N)}
    \, ,
    \label{eq:mesondecay}
\end{equation}
with $\lambda(1,x_\alpha,x_N) = (1-(x_N+x_\alpha)^2)(1-(x_N-x_\alpha)^2)$ and $x_i = m_i/m_m$.
The HNL decays via the weak interactions into final states that depend on the HNL mass $m_N$. We emphasise that the decay into three light neutrinos does not lead to a visible final state.
This invisible mode is the largely dominant decay mode for $m_N \leq 2m_e \sim 1$~MeV. 
For $m_N = {\cal O}(100)$~MeV and above, about 95\% of the decay modes include charged particles (or very quickly decaying neutral mesons) and thus produce visible final states.
The branching ratios for all possible final states were computed in \cite{Bondarenko:2018ptm} and we shall use their results in what follows.\footnote{We remark that the branching ratio of a given meson into an HNL is suppressed compared to the regular weak decay channels. Hence, for HNL production one considers mesons that decay almost exclusively via the weak interactions: additional decay modes from QED and QCD, unless additionally suppressed (by e.g. loops) yield large partial widths, that would render the meson branching fraction into an HNL phenomenologically irrelevant.}
The weak interactions also govern HNL decays into SM final states which include, due to the HNL mass being sub-GeV, neutrinos, leptons, and mesons.
The HNL partial decay widths depend in a non-trivial way on the mixing matrix elements $U_\alpha$ due to thresholds effects, and one can separate them into the following classes $|U_e|^2\Gamma_e,\,|U_\mu|^2\Gamma_\mu,\,|U_\tau|^2\Gamma_\tau$. 
We can define $\tau_\alpha = \hbar /\Gamma_\alpha$, which corresponds to the $N$ lifetime under the assumption that $U_\beta = 0$ for all $\beta \neq \alpha$.
The lifetimes $\tau_\alpha$ were evaluated for $\alpha=e,\mu,\tau$ and $|U_\alpha|^2=1$ in \cite{Bondarenko:2018ptm}, and we use these results to quantify the $N$ lifetime for generic mixing angles as follows:
\begin{equation}
    \tau = \left(\sum\limits_{\alpha=e,\mu,\tau}\frac{|U_\alpha|^2}{\tau_\alpha}\right)^{-1} \, .
\end{equation}
Its lifetime determines the decay rate of $N$ via the exponential decay law, which fixes the laboratory decay length $\lambda$,
\begin{equation}
    \lambda = c \tau \beta \gamma = c \tau  \frac{E}{m_N} \, ,
\end{equation}
which determines the travelled path of $N$ in its environment.
\subsection{Atmospheric LLP production}
Primary cosmic rays impinge on the Earth's atmosphere continuously and over a vast range of energies. Their interactions with atmospheric nuclei create cascades of light SM particles that turn into extensive air showers at energies above the TeV range.
In these showers charged and neutral mesons are produced, pions and kaons being the most copious among them, but above several tens of TeV the heavy-flavor mesons $D_0, D^+, D_s$ can also be produced.\footnote{We stress that heavy flavors are largely suppressed compared to lighter ones. In particular, it was shown in ref.~\cite{Arguelles:2019ziu} that the tau lepton flux is suppressed by about two orders of magnitude compared to the $D$ meson fluxes.}

To quantify the atmospheric spectra of different mesons we employ the tool \mceq~\cite{Fedynitch:2015zma}, which numerically assesses particle cascades in the Earth's atmosphere \cite{Fedynitch:2017trv}. This is achieved by solving the cascade equations that describe the evolution of particle densities as they propagate through a gaseous or dense medium. 
It is possible to select from various models for particle interactions and atmospheric density profiles.
Particles are represented by average densities on discrete energy bins, i.e.\ differential energy spectra, which we denote with $\Phi_i$, where $i$ is a label referring to a particle, and in the \mceq  output $\Phi_i$ is expressed in units of $({\rm s~cm^2~ sr~E)^{-1}}$.

We implement the decays of the mesons into HNL and charged leptons as new processes in \mceq according to eq.~\eqref{eq:mesondecay}, analogously to what was done in \cite{Arguelles:2019ziu}. The HNL flux is thus given by the summed decay spectra of all considered mesons,
\begin{equation}
    \Phi_N = \sum\limits_{m_i = \pi,K,D,D_s} \, \Phi_{m_i\to N}\,,
    \label{eq:PhiN}
\end{equation}
where $\Phi_{m_i\to N}$ is computed from \mceq and includes Br$(i\to N \ell)$ and the full kinematic information for $N$.
Notice that the spectra $\Phi_N$ depend on the zenith angle $\theta$, which is inherited from the spectra of the parent mesons.
We checked numerically that there there are no significant differences between the corresponding spectra for positively charged and negatively charged mesons in the energy ranges of our interest. We, therefore, decided to simulate only the spectra for positively charged mesons and to multiply our results by a factor of two, to account for mesons of both charges. 
The uncertainty corresponding to this is below the ${\cal O}(10)\%$ level.

For efficient generation, we set the lifetime $\tau_N$ in \mceq to infinity, and we add the effects from finite lifetimes by folding $\Phi_N$ with the decay probability
\begin{equation}
P_{decay}(d,l,\lambda) = e^{-\frac{d}{\lambda}}\left(1 - e^{-\frac{l}{\lambda}}\right) \, .
\label{eq:Pdecay}
\end{equation}
which defines the probability of $N$ with laboratory length $\lambda$ to decay after propagating a distance $d$ and before reaching the distance $d+l$.
To be explicit, we set the production height of the HNL to 20~km above the Earth's surface.
We have checked that this choice only affects our results quantitatively at the subpercent level, as expected since most of the distances between the considered detectors and the production point is typically at least ${\cal O}(1000)$~km. 

\subsection{Appearing Cherenkov Showers}
\begin{figure}
    \centering
    \includegraphics[width=0.45\textwidth, height=3.5cm]{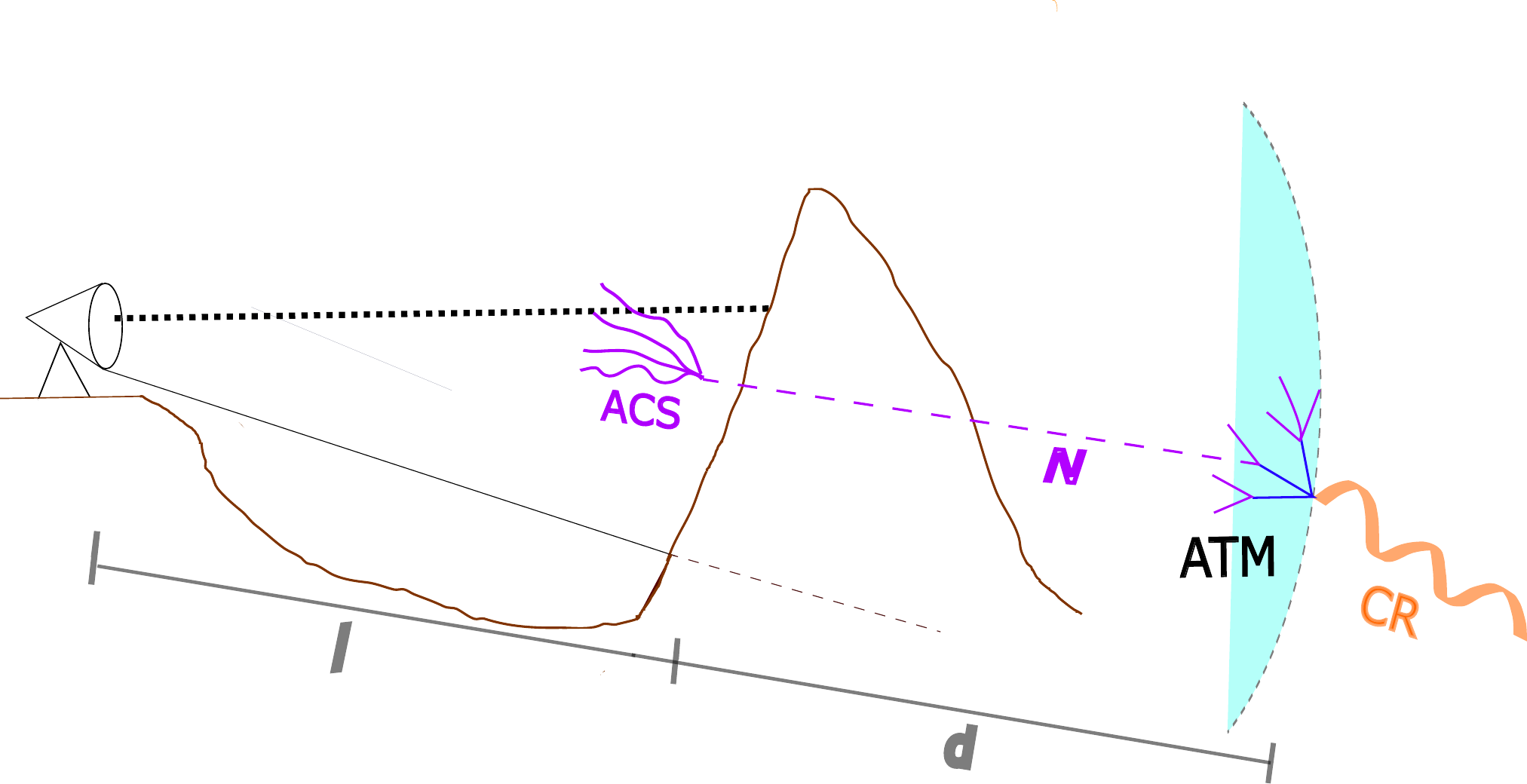}
        \includegraphics[width=0.45\textwidth, height=3.5cm]{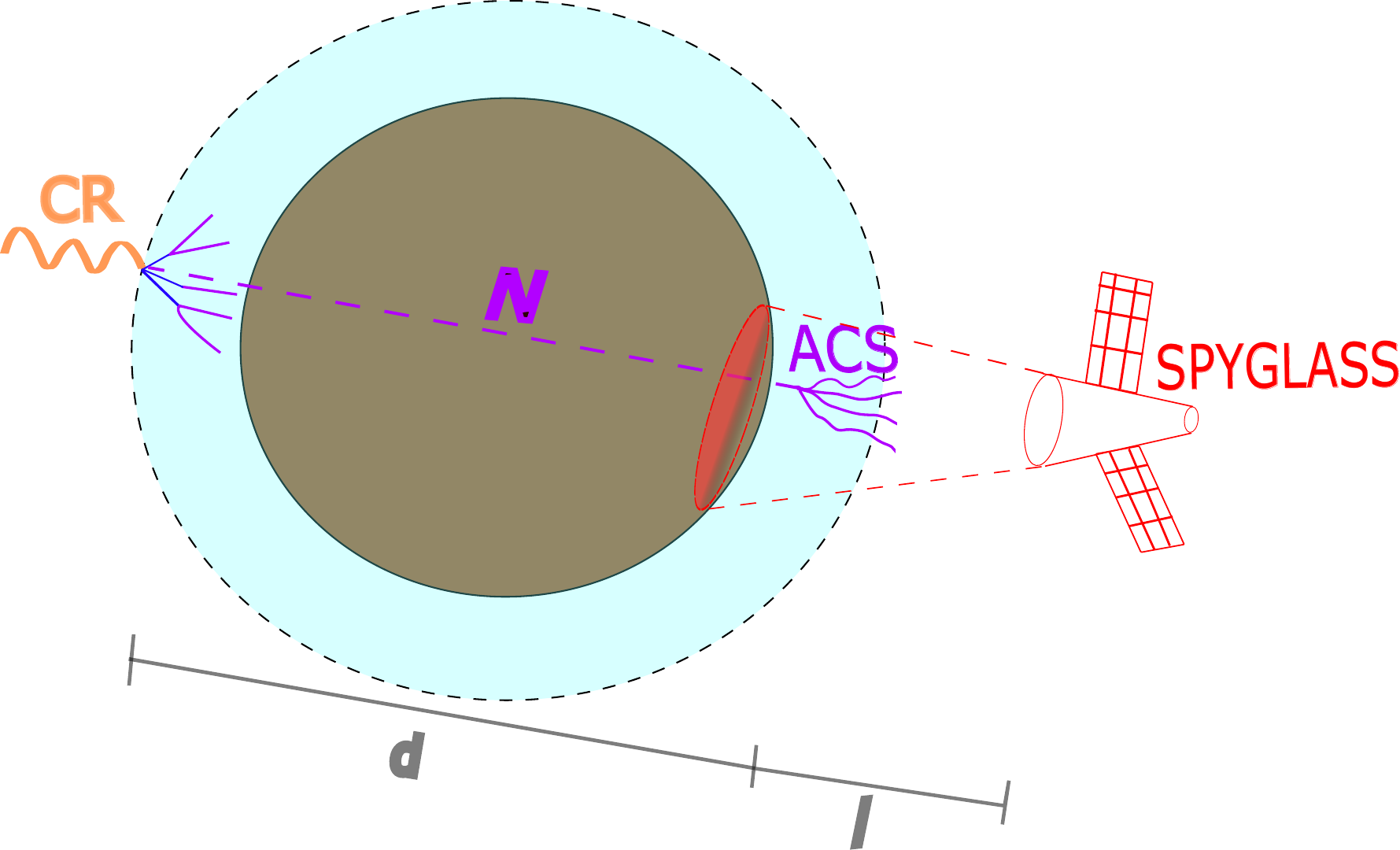}
    \caption{
    Left: a high energy cosmic ray (orange) impacts on the atmosphere (black line) generating a cascade of 
    secondary particles (purple). A long-lived neutral particle, here the heavy neutral lepton $N$, is produced in the atmosphere (shaded in cyan) at a distance $d$ from the mountain range (brown line), which is at a distance $l$ from the detector and acts as a shield. The $N$ decay into charged particles generates an ``Appearing Cherenkov Shower" (ACS, in purple) that is detected with a Cherenkov telescope operating in sub-horizon mode, with an angle $\alpha$ (blue lines). Right: A similar setup, but in this case the $N$ particle crosses the Earth's core and is observed by a satellite in geostationary orbit.  }
    \label{fig:geometry}
\end{figure}

All possible particles are produced in extensive air showers in the atmosphere at a distance $d$ from a shield, which can be a mountain cliff or Earth itself.
Due to their long lifetimes and their suppressed interactions, the produced $N$ particles propagate through the shield, to decay in the open air. Their decays into SM charged particles, which are highly boosted and move faster than the speed of light in air, produce a characteristic Cherenkov radiation.
The signatures we will discuss in the following are thus Cherenkov showers that appear in the air behind the shield, which we will call `Appearing Cherenkov Showers' or ACS for short. The above description is illustrated schematically in figure~\ref{fig:geometry} for an Earth-based telescope (left panel) and for an orbiting satellite (right panel).

The differential rate of ACS from visible $N$ decays in the volume behind the shield is a function of energy, time and the observed solid angle $\Omega$. It is given by 
\begin{equation}
    \frac{d^3R}{dEdt d \Omega}  = \text{Br}({N \to \rm vis})\,\Phi_N\, A\, P_{decay} (d,l, \lambda) \, ,
    \label{eq:diff_ACS_rate}
\end{equation}
where Br$({N \to  \rm vis})$ is the visible branching fraction of $N$ that we take from ref.~\cite{Bondarenko:2018ptm}, $\Phi_N$ is the HNL flux as in eq.~\eqref{eq:PhiN}, A is the observed area and the decay probability $P_{decay} (d,l,\lambda)$ is defined in eq.~\eqref{eq:Pdecay}.
The differential rate in eq.~\eqref{eq:diff_ACS_rate} lets us define the ACS spectrum by integrating out the time $t$ and observed solid angle $\Omega$,
\begin{equation}
    \frac{dR}{dE} = T\, \int_{\Omega_D} \frac{d^2R}{dE d\Omega} d \Omega\, ,
    \label{eq:dR/dE}
\end{equation}
where $T$ is the total observation time (fluxes are defined as averages so there is no explicit time dependence and $T$ can be integrated out trivially), $\Omega_D$ is the solid angle observed by the detector.
Likewise, a total count can be defined by integrating out the energy,
\begin{equation}
    R = \int_{E_1}^{E_2} \frac{dR}{dE} dE\, ,
\end{equation}
where $E_1,E_2$ are lower and upper energy thresholds, respectively, that are dependent on the considered detector and limit its sensitivity. As the detectors are typically efficient for any energy above threshold, and the spectrum behaves as a power law 
 $\sim E^{-3}$, we set $E_2 = \infty$ in practice.

\subsection{Backgrounds}
Most SM particles that are produced in cosmic ray induced atmospheric showers have short lifetimes and short interaction lengths and do not reach the Earth itself. Particles that arrive at the Earth's surface are mostly muons and neutrinos, with the latter penetrating a few meters up to a few kilometers into the ground, depending on their energy.

The energy of high-energy muons travelling through rock follows the equation (cf. chapter 34, ref~\cite{Workman:2022ynf}) %
\begin{equation}
    \frac{dE_\mu}{dX} = -\alpha - \beta E_\mu \qquad \text{with}\qquad \alpha = 2 \frac{\text{MeV}}{\text{g cm}^2},\, \quad \beta = 4 \cdot 10^{-6} \frac{1}{\text{g cm}^2} \, ,
\end{equation}
and using the density of ``standard rock'' of 2.65 g/cm$^2$ and $E_0 = 10$~TeV, this gives a radiation length of about 100~m.
We therefore consider this background irrelevant for shields that are at least ${\cal O}(1)$~km in thickness, which is the case for the experiments discussed below.

High energy neutrinos can interact with air or rock along the line of sight of an experiment, thereby creating charged or shortlived particles (like neutral pions) that in turn can give rise to a Cherenkov shower. To quantify the rates of this kind of background one needs to evaluate
\begin{equation}
    R_\nu(E_\nu) = \Phi_\nu(E_\nu) \, A \, T \, \sigma_\nu(E_\nu) \int_0^l  \rho(x) dx\,,
    \label{eq:nu_scattering}
\end{equation}
where $\Phi_\nu$ is the flux of all neutrino species through the observed surface $A$, $T$ is the observation time, $\sigma_\nu$ is the energy dependent neutrino scattering cross section, and $\rho(x)$ is the number density of scatterers along the path $l$.
For $l$ close to the horizontal, $\rho(x)$ can be taken constant, while for vertical $l$ one has to factor in the change of density.

Another possible source of background is artificial radiation produced by human activities. For instance, nuclear reactors and particle accelerators will radiate in appreciable amounts, acting as ``point sources''. They are not relevant for our purposes since a) their typical energies are well below the GeV scale and b) knowing their positions, they can be avoided by placing the detector far from the sources.

\section{Quantifying experimental sensitivities}
\label{s.sens}
In this section, we use the previously described methodology to quantify the rate of ACS from $N$ decays in the volume between a shield and an experimental detection device. 

First we consider two existing cosmic ray beam dump experiments, namely the Cherenkov telescope SHALON~\cite{Sinitsyna:2020kco} and the Antarctic Impulsive Transient Antenna (ANITA)~\cite{ANITA:2010ect}, which use a mountain and the whole of the Earth as a shield, respectively. We evaluate their sensitivity to HNLs and we conclude that they are not competitive with existing constraints, and near-future planned experiments. 
Then we use the SHALON and ANITA experiments discussed above as baseline for two dedicated HNL detectors with optimised design parameters: a ground-based telescope and a space-based experiment. 

We collect the relevant experimental design parameters (geometrical, exposure time, energy threshold) in Table~\ref{tab:experiments}.
\begin{table}
    \centering
    \begin{tabular}{c|cccccc}
         & d [km] & l [km] & A [cm$^2$] & solid angle [sr] & T [s] & $E_1$ [GeV] \\
         \hline\hline
    SHALON     & 1520 & 7 & $7\times 10^9$ & 0.001 & 1166400 & 800 \\
    SHALON 10     & 1520 & 7 & $7\times 10^9$ & 0.001 & $3.15\times 10^7$ & 10\\
    ANITA      & 7000 & 37 & $4 \times 10^{12}$ & $2\pi$ & 5011200 & $10^8$  \\
    MtThor     & 300 & 5.67 & $10^{10}$ & 0.024 & $3.15\times 10^8$ & 10\\
    SPYGLASS X    & 12800 & 35786 & $10^{16}$ & $8\times 10^{-4}$ & $3.15\times 10^7$ & \small{X=}{10, 100, 1000} \\
    \end{tabular}
    \caption{Detector specifications considered in this article: original SHALON~\cite{Sinitsyna:2020kco} and ANITA~\cite{ANITA:2010ect} designs, and the three modifications: a) a SHALON-like experiment with a 10 GeV energy threshold (dubbed SHALON-10), b) an improved SHALON experiment taking as a reference Mt Thor, in Baffin Island, Canada (MtThor) and c) an improved ``Anita'' setup, which we call SPYGLASS (Satellite PreYing on Geo Long-lived Appearing ShowerS). The choice of optimal parameters is discussed in this section. Here $d$ is the distance between shield and HNL production in the atmosphere, $l$ is the distance between detector and shield, $A$ is the detector-observed surface, the solid angle as subtended by the experiment, $T$ the observation time, and $E_1$ the lower energy threshold.}
    \label{tab:experiments}
\end{table}

\subsection{SHALON} 
SHALON is an imaging atmospheric Cherenkov telescope\footnote{We note that many existing gamma-ray telescopes, like HESS~\cite{Bernlohr:2003tfz,Cornils:2003ve}, MAGIC~\cite{Baixeras:2004hd} and VERITAS~\cite{https://doi.org/10.48550/arxiv.0912.3841} observe air showers directly, without any shielding, which makes them insensitive to LLP signals in ACS.} located at Tien-Shan High Mountain Station, equipped to record precise information about atmospheric shower structures at the very high energies \cite{Sinitsyna:2020kco}. Concretely, its base design aims at studying gamma-rays in the 0.8 - 100 TeV energy range.
It recorded five extensive air showers coming from an opposite slope of a gorge in 7~km distance over a period of 324 hours at sub-horizon zenith angles at $\theta = 7^\circ$ with energies of \{11, 7, 6, 8, 17\} TeV.
A relevant source of background is charged current interactions of high-energy neutrinos with the rock. These interactions can produce high-energy muons or leptons which subsequently decay in air to produce an ACS. The collaboration excluded this possible background as explanation for the observed ACS due to the limits on the neutrino fluxes for energies around 10~TeV \cite{Sinitsyna:2009dn}.
We want to inspect briefly the possibility that these showers stem from HNL decays as was conjectured by the collaboration~\cite{Sinitsyna:2009dn,Sinitsyna:2013hmn}.  The ACS spectra from HNL decay products is given by eq.~\eqref{eq:dR/dE} and, using that $\Phi_N \propto \Phi_i \propto E^{-3}$ (where we neglect the fluxes' deviations from the power law, which are small for the considered energy range, cf.\ the dashed lines in figure~\ref{fig:spectra_examples} below) we simplify the energy dependence of $dR/dE$ and write:
\begin{equation}
    \frac{dR}{dE} \propto \frac{P_{decay}}{E^3} \, .
    \label{eq:LLPspec}
\end{equation}
This spectrum has a maximum where the laboratory lifetime $\lambda = c \tau E/m$ is comparable to the distance $d$, which needs to be around the observed energies of $E=10$~TeV.\footnote{Decays that occur closer to the telescope will likely lead to underestimated total energy of the shower~\cite{Sinitsyna:2009dn}.} 
The observed shower energies therefore put a condition on the proper lifetime:
\begin{equation}
    \left( \frac{d\, m}{\lambda c \tau E^2} - \frac{3}{E} \right) = 0 
    \qquad \Rightarrow \qquad 
    c \tau = \frac{d m}{3 E} \, ,
    \label{eq:lambda}
\end{equation}
where we have used that the length $l$ is much smaller than the travelled $d$ distance, $l \ll d$.
With eq.~\eqref{eq:lambda} we can estimate a range of proper lifetimes that matches the observed spectral shape.
Considering the range of observed event energies, 6~TeV to 17~TeV, and the SHALON specific distance of $d = 1520$~km (cf.\ tab.~\ref{tab:experiments}) we find for the exemplary mass of $m=0.1$~GeV that $\lambda_N \in (2.2, 6.3)$~m.

To compute the HNL-induced ACS spectrum for SHALON we use eq.~\eqref{eq:dR/dE} with the zenith angle $\theta = 97^\circ$ and the fact that the flux does not vary much within the telescope's field of view, such that the ACS energy spectrum can be approximated by:
\begin{equation}
    \frac{d R_{\rm SHALON}}{dE} = T\, \Omega_D\,\text{Br}({N \to \rm vis})\, \phi_N\, A\, P_{decay} (d,l, \lambda)\,,
\label{eq:R_SHALON}
\end{equation}
where the parameters $T,\Omega_D, d,l$ are listed in tab.~\ref{tab:experiments}, and $\phi_N$ is the $\Phi_N$ HNL flux from equation \eqref{eq:PhiN} evaluated at a zenith angle of $\theta = 97^\circ$.

To illustrate the main features of this spectrum we fix the decay length to $\lambda = 5 \cdot(m_N/0.1~\text{GeV})$~m and show the computed spectra for different HNL masses in figure~\ref{fig:spectra_examples}, considering only the electron final state  $|U_\alpha|=\delta_{\alpha e}$.
The dashed lines denote the total flux of $N$ while the solid lines denote the flux of $N$ decaying in the SHALON observable decay volume.

We make five observations: i) the contributions from pions and kaons are roughly of the same magnitude; ii) contributions from $D_s$ mesons are suppressed by about two orders of magnitude and including Br$(D_s \to N \ell_e) = {\cal O}(10)\%$ further increase this suppression; iii) all contributions exhibit their maxima around 10~TeV as designed; iv) when the proper lifetime $c \tau$ varies, longer (shorter) lifetimes shift the maximum to smaller (larger) energies, with the maximum flux increasing (decreasing) in magnitude; v) the limited decay volume acts as a ``band-pass'' filter, probing energies in a window around the desired peak, with the window width depending on the parent particle and the HNL mass.
\begin{figure}
    \includegraphics[width=0.32\textwidth]{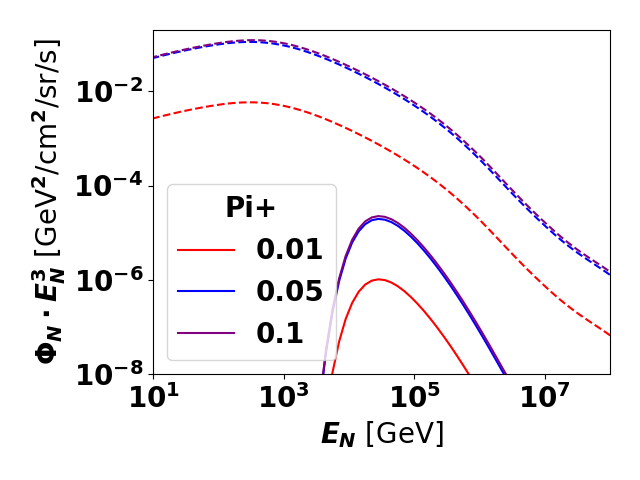}
    \includegraphics[width=0.32\textwidth]{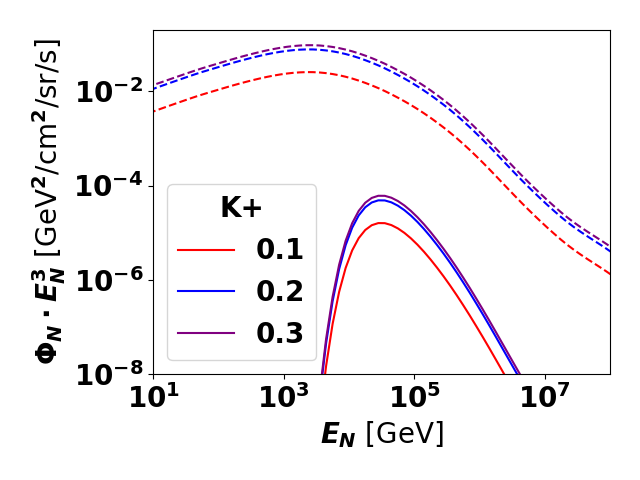}
    \includegraphics[width=0.32\textwidth]{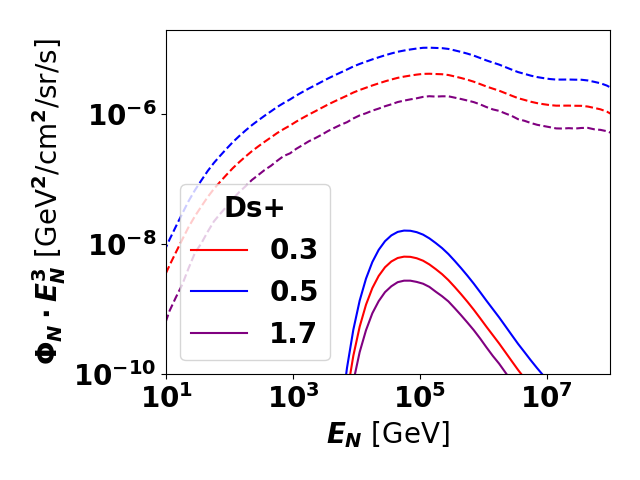}
    \caption{Contributions to the atmospherically produced HNL fluxes $\Phi_N \cdot E_N^3$ from pion (left panel), kaon (middle panel) and $D_s$ meson (right panel) in units GeV$^2/(s\, cm^2\,sr)$. The contributions from the different mesons to the HNL fluxes are shown for different HNL masses in GeV as indicated in the legends of each panel.
    Dashed lines denote the \mceq produced fluxes for $\tau_N =\infty$, while solid lines include the decay factor in eq.~\eqref{eq:Pdecay}, and using the distance $d=1520$~km and decay volume length $l=7$~km for the SHALON experiment.
    In this figure we fixed $|U_{e}| = 1$ and $|U_\alpha|^2 = 0$ for $\alpha=\mu,\tau$.}
    \label{fig:spectra_examples}
\end{figure}

We find that the event rate observed by SHALON cannot be reproduced for the production processes of $\pi^+ \to N \ell^+$ or $K\to N \ell$ for branching ratios that are not already excluded by experimental data.
In particular, even for laboratory lifetimes around 5~m the observed event rate $R_{\rm SHALON} = {\cal O}(1)$ between between 6 and 20~TeV can be matched only for BR$({i \to  N \ell}) = {\cal O}(1)$, with $i=\pi,\,K$. This can be interpreted as a sensitivity to the active-sterile mixing parameter $|U_\alpha|^2$ which is completely superseded by other experiments~\cite{Abdullahi:2022jlv}. 
SHALON's sensitivity to HNL could be increased significantly if the energy threshold were to be lowered. 
We evaluate the sensitivity for a hypothetical new run of this experiment, considering an energy threshold of 10~GeV~\footnote{The MAGIC Cherenkov Telescope lowered its original 50 GeV~\cite{Baixeras:2004hd} threshold down to 25 GeV during operations~\cite{MAGIC:2008jib}.  } instead of 800~GeV and an observation time of a whole year. The parameters are shown in tab.~\ref{tab:experiments}.
Assuming that the experiment remains free of backgrounds for the new energy thresholds and considering that the event rate $R$ follows a Poisson distribution, a non-observation of $3$ predicted events excludes a given parameter space point at 95\% confidence level.

\begin{figure}
    \centering
    \includegraphics[width=0.49\textwidth]{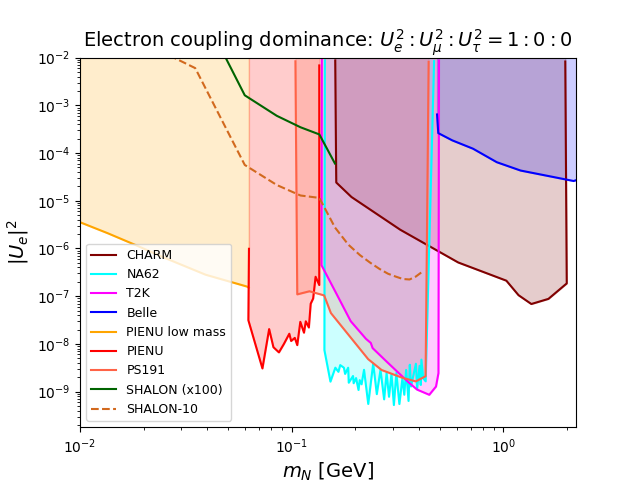}
    \includegraphics[width=0.49\textwidth]{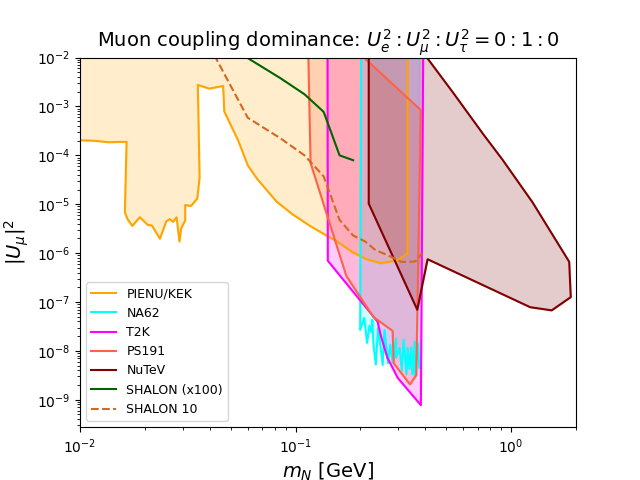}
    \caption{Sensitivity of the SHALON Cherenkov Telescope to Heavy Neutral Leptons coupling to the first (left panel) and second (right panel) lepton generation, confronted with current experimental constraints. Shadowed areas show the current experimental constraints from PIENU~\cite{PIENU:2017wbj} peak (orange), and low-mass interpretation searches~ \cite{PiENu:2015seu,Bryman:2019ssi,Bryman:2019bjg} (red), PS191~\cite{Bernardi:1987ek} (light red), NA62~\cite{NA62:2020mcv,NA62:2021bji} (cyan), T2K~\cite{T2K:2019jwa} (pink), CHARM~\cite{CHARM:1985anb} (brown) and Belle~\cite{Belle:2013ytx} (blue). The solid green line shows the expected sensitivity of SHALON with its current dataset multiplied by a factor 100, for better visualization, while the dashed light-brown line shows the expected reach of a similar experiment with a 10 GeV threshold and 10 years of exposure.
    }
    \label{fig:SHALONplot}
\end{figure}

The resulting sensitivity is shown in figure~\ref{fig:SHALONplot} in a solid dark green line (for visualization purposes we plotted the expected sensitivity multiplied by a factor of 100), and the lower-threshold, larger exposure variant SHALON 10 is shown in a light-brown dashed line. Colored shadowed areas indicate the current existing experimental constraints (details in caption), taken from~\cite{Abdullahi:2022jlv}. 
The figure shows that even with the lower energy threshold and increased observation time, SHALON would not be sensitive to untested parts of the HNL parameter space. We emphasize that the observed ACS by SHALON cannot be explained via the decays of atmospherically produced LLPs in general, unless their flux is comparable to the fluxes of light mesons. Insisting on such an explanation would require a new production mechanism, which necessarily implies an extension of our HNL model.

\subsection{ANITA}
\begin{figure}
    \includegraphics[width=0.32\textwidth]{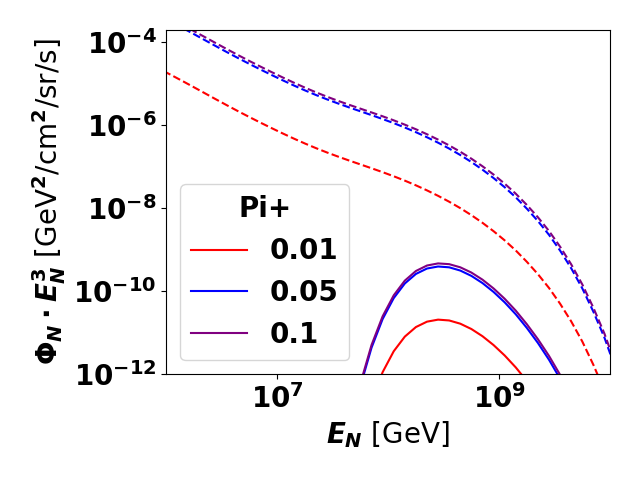}
    \includegraphics[width=0.32\textwidth]{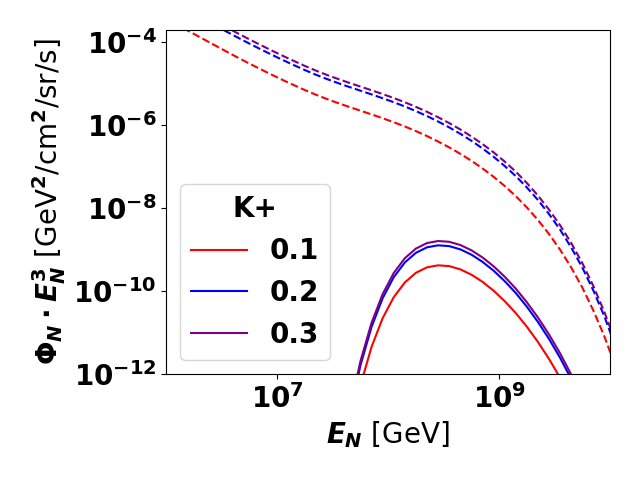}
    \includegraphics[width=0.32\textwidth]{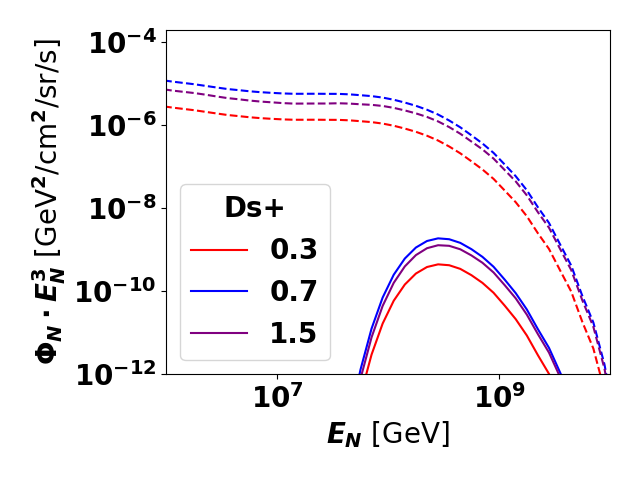}
    \caption{Analogous to figure~\ref{fig:spectra_examples}, for the ANITA experiment with distance $d=7000$~km and decay volume length $l = 37$~km.}
    \label{fig:spectra_ANITA}
\end{figure}
ANITA is a stratospheric balloon-borne experiment flying at a height of about 37,000 meters and its four flights set the most competitive limits on the ultrahigh-energy (UHE) diffuse neutrino flux above several tens of EeV~\cite{ANITA:2010ect}.
It uses an array of radio antennas to detect the impulsive radio geosynchrotron emission from UHE cosmic ray induced air showers, in particular those from 
UHE neutrino interactions with the Antarctic ice sheet. 

Being suspended 37~km above the ground, ANITA views the arctic ice sheet up to the horizon with a geometric area that approaches $400$~km$^2$ \cite{ANITA:2021xxh} and covering a solid angle close to $2\pi$~sr. 
The experiment detected two Earth-emergent UHE showers during its first \cite{ANITA:2016vrp} and third flight \cite{ANITA:2020sds},  corresponding to observation times of of 35 and 23 days, respectively.
Possible SM backgrounds are tau-induced Atmospheric Showers from Earth-skimming tau neutrinos, with well matching predictions for spectral properties and locations, however, the rates are far too high compared with existing limits on neutrino fluxes \cite{ANITA:2021xxh}.
Consequently these ACS cannot not be explained with known physics, and it was argued that they are strong hint for physics beyond the SM, see e.g.~\cite{Fox:2018syq,Collins:2018jpg,Hooper:2019ytr,Cline:2019snp,Heurtier:2019rkz,Borah:2019ciw,Anchordoqui:2019utb,Abdullah:2019ofw,Esmaili:2019pcy,Altmannshofer:2020axr,BhupalDev:2020zcy,Liang:2021rnv}.

We now discuss the possibility that these anomalous events are due to HNL decays, which can take place in air or in ice. The charged final state particles can give rise to geosynchrotron emissions that we expect should be very similar to those from neutrino interactions. 
To evaluate ANITA's sensitivity to HNL decays we set the distance between HNL production and decay equal to the larger chord length of the two Earth emergent events, $d \sim 10^7$~m, while for the length of the decay volume we use ANITA's altitude, $l = 37\times 10^3$~m.
A signal must not be exponentially suppressed, such that $\lambda$ and $d$ must be similar, which implies
\begin{equation}
 \lambda \gtrsim d\,  \frac{ m}{E} = 10^{-2} \text{m} \Bigl( \frac{m}{\text{GeV}} \Bigr) \Bigl( \frac{d}{10^7 \text{m}} \Bigr)  \Bigl( \frac{10^{18} \text{GeV}}{ E } \Bigr) \, ,
 \label{eq:ANITA_lifetime}
\end{equation}
This points towards relatively large lifetimes: for e.g.\ $m = 1$~GeV the decay length has to be $\lambda \gtrsim 0.01$~m. We remark that such longish lifetimes are somewhat disfavoured: for $l \lesssim d, \lambda$ then $P_{decay} \propto l / \lambda$ (eq.~\eqref{eq:Pdecay}), and hence the signal rate is inversely proportional to $\lambda$. The decay lengths required in eq.~\eqref{eq:ANITA_lifetime} are short as far as HNL are concerned, indicating masses above the GeV, where the production mechanisms are limited to heavy meson decays.

To compute the HNL-induced ACS spectra for ANITA one has to integrate the solid angle below the horizon. 
We obtain an approximation for the spectra by assuming that the flux is independent of the zenith angle, evaluating the flux $\Phi_N$ at a zenith angle close to the horizon.
Since the fluxes at high energies are larger for zenith angles closer to the horizon, our approximation can be interpreted as an upper bound on the expected ACS rate.
We checked numerically that this leads to an overestimation of the rate by an ${\cal O}(1)$ factor. The corresponding ACS energy spectrum can be evaluated via eq.~\eqref{eq:dR/dE} with the parameters $T,\Omega_D, d,l$ listed in tab.~\ref{tab:experiments} under the label ``ANITA''.

We show the flux of HNL decays in the ANITA observed volume 
in figure~\ref{fig:spectra_ANITA}, where we set the decay length $\lambda$ according to eq.~\eqref{eq:ANITA_lifetime} and we set $|U_\alpha| = \delta_{\alpha e}$. 
It is very clear that the upper limit on the event rates $R_{\rm ANITA}\ll 1$ for 0.02~GeV$\leq m_N\leq 1.8$~GeV.
This indicates that the Earth-emergent UHE events are not due to ACS from atmospherically produced HNLs.

\subsection{Cherenkov telescope at Mount Thor}
Having examined the sensitivity of SHALON and understanding its limitations, we now consider the possibility to build a SHALON-like experiment that is dedicated to test ACS from atmospherically produced HNL.
We therefore consider, for concreteness, the cliff of the Thor Peak on Baffin Island, Canada, as a natural location as it features Earth's greatest vertical drop of 1,250 m.\footnote{Many other places on Earth exist that may exhibit similar merit for such an experiment. However, it is beyond the scope of this article provide an exhaustive list of these possibilities.} We propose a Cherenkov detector with 
a solid angle coverage of $0.024$~sr (corresponding to an opening angle of 10$^\circ$, similar to the SHALON telescope)
that observes the cliff of the Thor Peak, the height of which together with the solid angle coverage determines the length of the decay volume, i.e.\ the distance between the cliff and the detector, as $5.67$~km.
We consider the detector to be built at ground level and to point at 5$^\circ$ above the horizon, which fixes the distance between HNL production in the atmosphere and the mountain peak to $\sim 300$~km. 
We remark that observation angles above the horizon affect the flux of high-energy muons that penetrate the upper parts of the cliff and enter the detection volume.\footnote{Observation angles where the detector points below the horizon increase the shielding from penetrating muons. However, pointing the detector below the horizon while focusing it on the cliff implies it must be built at higher altitudes, which in turn limits the possibilities for both its location and its deployment.}
This effect strongly depends on the topology of the cliff and its calculation is beyond the scope of this article. In the following, we will assume that the cliff is sufficiently massive to justify our assumption of being free of background.
Lastly, we assume that the detector can be technically realised with a lower energy threshold of $E_1=10$~GeV and that this dedicated experiment will observe the cliff for a time of 10 years. The experimental setup is identical to the left panel of figure~\ref{fig:geometry}.

Regarding the neutrino interaction backgrounds the experiment parameters are quite similar to those of SHALON, therefore we expect that the conclusions of a background-free environment from ref.~\cite{Sinitsyna:2009dn} hold.  
Under the assumptions of zero background, we again posit that a non-observation of $3$ predicted events excludes a given parameter space point at 95\% confidence. 

\begin{figure}
    \centering
    \includegraphics[width=0.49\textwidth]{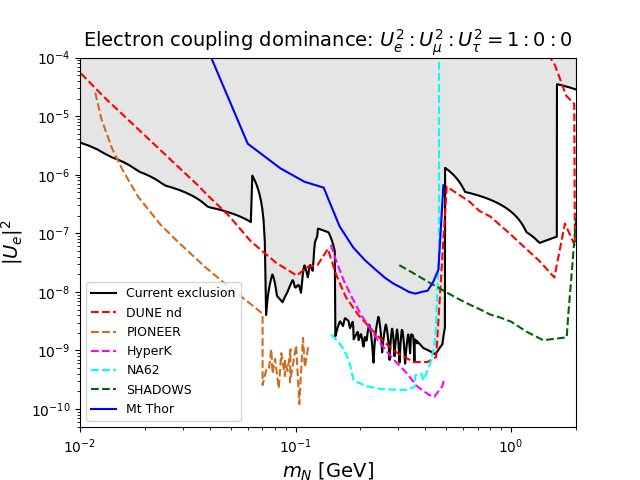}
        \includegraphics[width=0.49\textwidth]{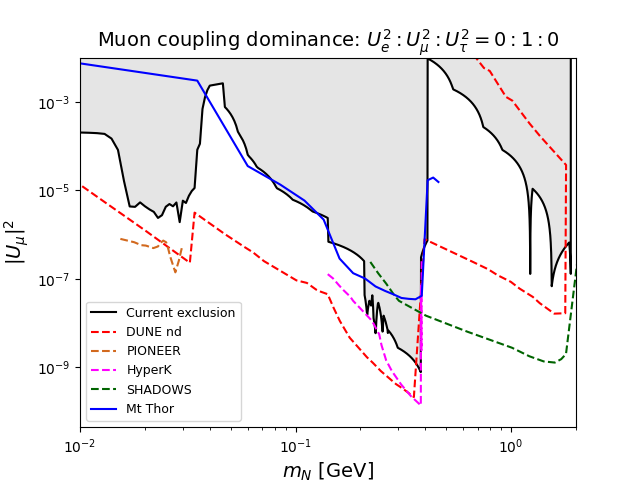}
    \caption{Sensitivity of the proposed Cherenkov Telescope at Mount Thor to Heavy Neutral Leptons coupling to the first (left panel) and second (right panel) lepton generation, confronted with current and future projected experimental constraints. Current constraints (already shown in figure~\ref{fig:SHALONplot}) are indicated in the grey shadowed area with a black boundary. Dashed lines indicate future projections for the DUNE near detector~\cite{Breitbach:2021gvv} (red), PIONEER~\cite{PIONEER:2022yag} (light brown), Hyper-K~\cite{T2K:2019jwa} (pink), NA62 ~\cite{NA62:2020mcv,NA62:2021bji} (cyan) and SHADOWS~\cite{Baldini:2021hfw} (dark green) experiments. The solid blue line shows the expected sensitivity of our Mount Thor experiment.
    }
    \label{fig:MtThorplot}
\end{figure}

The HNL-induced ACS spectrum observable at MtThor is evaluated following eq.~\eqref{eq:dR/dE}, with the experimental parameters as above summarised in tab.~\ref{tab:experiments} under the label ``MtThor''.
The corresponding sensitivity to ACS from atmospherically produced HNL decays is shown by the blue line in figure~\ref{fig:MtThorplot}. There we confront this sensitivity with existing constraints (black), and the expected reach of several experiments in the near future in dashed lines (see caption for details), taken from~\cite{Abdullahi:2022jlv}. 
The figure shows that the sensitivity of the MtThor experiment reaches into untested parts of the HNL parameter space of $U_\mu$ for masses above $\sim 40$~MeV and up to $\sim 400$~MeV.
Due to the suppression of the charmed meson fluxes the HNL flux is insufficient for $m_N > m_K - m_\ell$ to produce a signal rate of ${\cal O}(1)$ or higher, thus limiting the sensitivity.
We notice that the experiment exposure ($l\times A\times T$) needs to be enhanced by a factor $\sim 100$ in order to lead to a visible signal in the mass range $m_N > m_K$, which would incidentally render the sensitivity for $m_N < m_K$ comparable to the planned DUNE near detector~\cite{Breitbach:2021gvv}, and also for the anticipated PIONEER~\cite{PIONEER:2022yag} and Hyper-K~\cite{T2K:2019jwa} experiments.

\subsection{Particle detector in geostationary orbit}
We observe that the air-borne experiment ANITA benefits from a much larger area of observation and decay volume compared to ground-based experiments.
We therefore want to propose an experiment that maximises this effect by removing it `as far as possible' from the surface of the Earth, observing a very large area on the ground.

To be explicit, we propose to position a detector in a geostationary orbit above the equator at an altitude of about 35786 km, monitoring an area on the surface that has no background activity. 
We assume that the Sahara desert is an appropriate area for this purpose,\footnote{Or any other large deserted area, with little human activity. The choice of a specific area is beyond the scope of this work.} and that our detector should view a fraction of about $10^6$~km$^2$ of it, which corresponds to a solid angle of $8\times 10^{-4}$~sr. Since the experimental setup reminds us of a spotting apparel, we tentatively dub our experiment SPYGLASS (Satellite PreYing on Geo Long-lived Appearing ShowerS).

HNL that are produced on the far side of the earth would travel through the whole of Earth and decay within the $\sim 35000$~km between the Earth and the experiment. 
For LLP with lifetimes that enable their propagation over $10^4$ km, most of their decays take place in vacuum.
Hence as signal we consider appearing charged particles --the daughters from the HNL decay in vacuum-- rather than Cherenkov light. 
Therefore, a suitable experiment would be similar to other space-borne particle detectors, like for instance the PAMELA apparatus, which comprises several detector components in order to reliably identify antiparticles from a large background of other charged particles \cite{Picozza:2006nm} with energies up to 200~GeV, or the DAMPE experiment, which can detect and identify particles up to and beyond 10~TeV \cite{DAMPE:2017cev}.
The detailed discussion of technical details is beyond the scope of this article.

\begin{figure}
    \centering
    \includegraphics[width=0.49\textwidth]{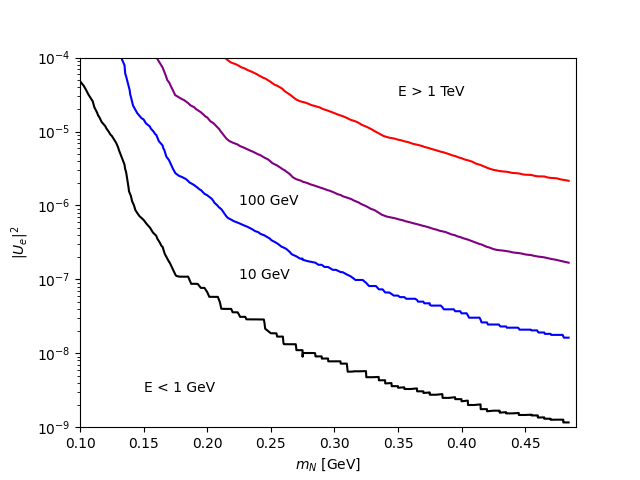}
    \includegraphics[width=0.49\textwidth]{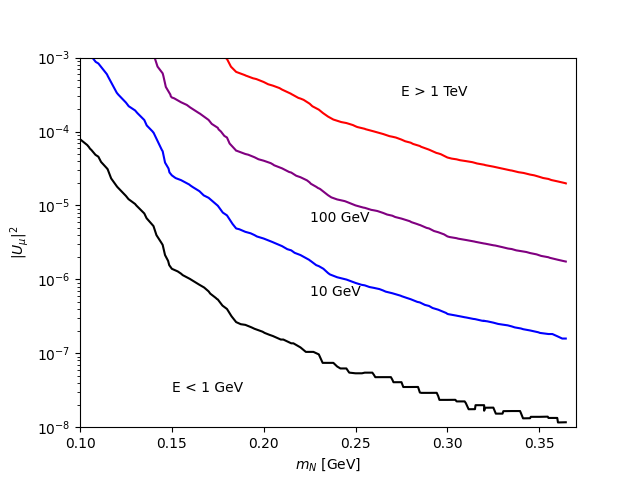}
    \caption{Contours of average energy of $\Phi_N$ as defined in eq.~\eqref{eq:average_energy} for HNL mixing exclusively with the electron (left) or muon flavor (right), evaluated at a zenith angle close to the horizon.}
    \label{fig:average_energy}
\end{figure}

The experimental setup is similar to ANITA (cf.\ \ref{fig:geometry}). An HNL would need to propagate through the whole Earth and thus have very long laboratory lifetimes, $c \tau \sim {\cal O} (10$ km). Depending on the proper lifetime, the average energy
\begin{equation}
    \overline E = \frac{\int E \Phi_N dE}{\int \Phi_N dE} \, ,
    \label{eq:average_energy}
\end{equation}
of the ACS energy spectrum can vary between a few GeV and well above the PeV scale (and depends on the zenith angle). To illustrate this point, we show the contours of average energies in the mass-mixing parameter space in figure~\ref{fig:average_energy}, where a zenith angle close to the horizon was used.

We observe that the contour lines follow those of constant lifetimes, which is due to the fact that a fixed lifetime determines the boost factor and thus the energy that is required for $N$ to propagate through the entire Earth.
Naturally, for parts of the parameter space where the proper lifetime is ${\cal O}(1)$~m or larger (for $m_N \sim 0.1$~GeV and $|U_\alpha|^2 < 10^{-5}$) no substantial boost is required, such that the average energy is close to that of the parent spectrum, which prefers smallest possible energies. 
Therefore, to probe smaller masses and mixings one requires lower detection thresholds.
The signal's energy is a crucial parameter in controlling possible backgrounds.
Background rates from neutrino scattering in the Earth's atmosphere can be estimated using eq.~\eqref{eq:nu_scattering} with the neutrino cross section (PDG) of $0.8\,E_\nu\, 10^{-38}$ cm$^2/$GeV and the dominant muon neutrino flux of $\Phi_\nu(E_\nu) = 40~({\rm m^2~s ~sr})^{-1} (10\text{ GeV}/E_\nu)^3$. We then expect about $\sim 100$ scattering events for $E_\nu = 10$~GeV, and less than 1 event for $E_\nu \geq 100$~GeV per year over the portion of the desert observed by SPYGLASS.
These scattering events produce Cherenkov light in the atmosphere, which the experiment can exploit to veto this background.
Other conceivable backgrounds at this energy are cosmic rays, which can only enter SPYGLASS when they are reflected off the Earth. We neglect this background and consider the signal of appearing charged particles with energies above a certain threshold as free of background.

\begin{figure}
    \centering
    \includegraphics[width=0.49\textwidth]{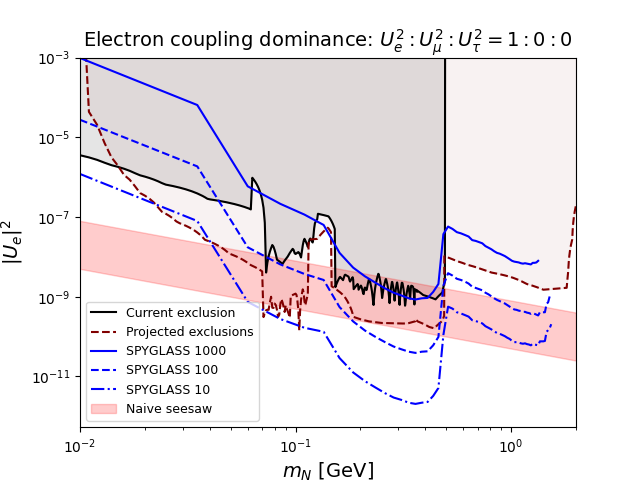}
    \includegraphics[width=0.49\textwidth]{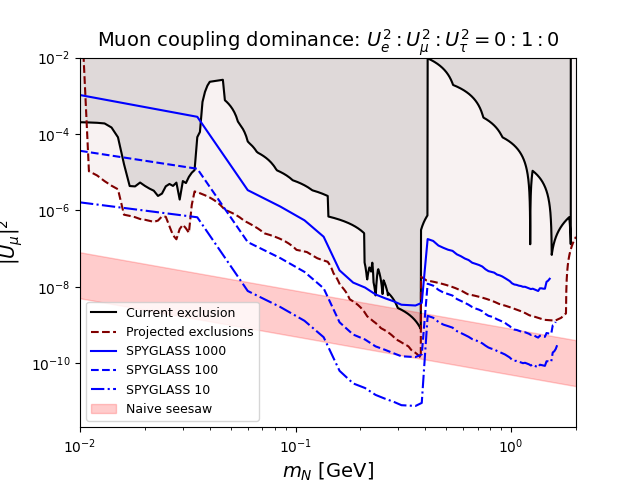}
    \caption{
    Constraints on the parameter space of a Heavy Neutral Lepton coupling to the first lepton generation (left panel) and second lepton generation (right panel) when considering the particle detector satellite SPYGLASS. The shadowed grey (brown) area with a solid black (dashed brown) boundary is (will be) excluded by current (future) constraints, already shown in detail in figure~\ref{fig:SHALONplot} (\ref{fig:MtThorplot}). The blue lines indicate the expected sensitivity of SPYGLASS for energy thresholds of 1000 GeV (solid), 100 GeV (dashed) and 10 GeV (dot-dashed). The red band is obtained from the na\"ive seesaw expectation for the mixing angle (equation ~\ref{eq:seesaw}) by varying the heaviest SM neutrino mass $m_{\nu_3}$ between the minimal allowed value (by setting the lightest neutrino mass to zero) and the laboratory bound of 0.8 eV~\cite{KATRIN:2019yun}. 
    }
    \label{fig:resultsGeo}
\end{figure}

The HNL-induced ACS rate for the satellite experiment is evaluated analogously to eq.~\eqref{eq:R_SHALON}, with the experimental parameters as above summarised in tab.~\ref{tab:experiments} under the label ``SPYGLASS'', and using a zenith angle close to the horizon. 
To be explicit, we show results for three different energy thresholds, namely 10, 100, and 1000~GeV, which are displayed in blue dot-dashed, dashed and solid lines in figure~\ref{fig:resultsGeo}. The above-mentioned experimental details are listed in tab.~\ref{tab:experiments}. These projections are compared vis-\`a-vis the existing constraints from figure~\ref{fig:SHALONplot} (gray area) and the future projections from in figure~\ref{fig:MtThorplot} (brown area). We also included a red band,
corresponding to the na\"ive seesaw relation, equation~\ref{eq:seesaw}, for $m_{\nu} = 0.8 $ eV (direct KATRIN
limit~\cite{KATRIN:2019yun,KATRIN:2021uub})  and for $m_{\nu} = 0.05$ eV. The latter value corresponds to the lowest possible mass for third neutrino, enforcing the 
lightest neutrino to be massless (in normal ordering).

We conclude that both SPYGLASS 10 and 100 would largely extend the coverage of the HNL parameter space, well beyond the reach of planned experiments and largely probing the region na\"ively disfavoured by BBN consideration, which is always taken as an indicative bound in the literature~\cite{Abdullahi:2022jlv}. SPYGLASS 10 will, in addition, go well below the na\"ive see-saw band, probing neutrino mixing parameters $|U_\alpha|^2$ below $10^{-11}$. It is therefore interesting to further explore the physics opportunities that such an experimental setup would open.~\footnote{It is clear that this setup offers the largest possible shield on planet Earth, where we have seized the knowledge of meson production in the atmosphere accumulated over many decades of cosmic ray experiments. Considering an equivalent setup on another planet is an academically appealing option, which would require a careful understanding of the atmosphere of the target planet. }.

\section{Conclusions}
\label{s.conclu}
In this work we discussed that Long-Lived Particles can be tested in dedicated ``cosmic-ray beam dump'' experiments, being produced in extensive air showers in the atmosphere where the Earth acts as the beam dump for the SM particles produced in the shower (except for neutrinos) thus providing a new laboratory that is complementary to searches at collider and beam dump experiments. Key to testing atmospherically produced LLP is detecting the LLP decay products in existing or dedicated new detectors together with the fact that propagation of SM particles through the Earth is limited.

We have examined the sensitivity of two experiments: the Earth-based Cherenkov telescope SHALON and the balloon-borne radio antenna detector ANITA, the selection being motivated by intriguing excesses in the datasets of both experiments, with claims in the literature of a possible Beyond the Standard Model origin. 
We have employed a toy model of Heavy Neutral Leptons, which is  from the class of Standard Model extensions that can account for the origin of neutrino masses.
This model can feature sub-GeV neutral particles with laboratory lifetimes of several kilometers, thus promising a suitable explanation for the observed excesses. We have considered the cases where the HNL is predominantly coupled to either the first or the second generation of leptons.

We have verified for both experiments that the respective anomalies can not be explained with HNL, as the sensitivity of these experiments in the relevant mass range (10 MeV - 2 GeV) is below that of existing constraints. 
Using the SHALON and ANITA designs as baselines we then proposed two new experiments with optimised design for HNL detection: a) ``MtThor'', a SHALON-like telescope located in a mountain cliff with a large vertical descent, largely improving the solid angle covered by the telescope; b) SPYGLASS (Satellite PreYing on Geo Long-lived Appearing ShowerS), a satellite in geostationary orbit scanning a portion of the Earth surface without man-made radiation, which for concreteness we consider to be part of the Sahara desert. SPYGLASS' technical capabilities (energy thresholds, particle identification, etc) are inspired by the PAMELA satellite. 

We have found that the proposed experiment at Mount Thor can test currently untested parts of the HNL parameter space for HNL with masses in the range of 10-500 MeV, i.e.\ below the kaon mass.
It is worthwhile pointing out that enhancing the exposure of the experiment by a factor $\sim 100$ would lead to a sensitivity that is at least comparable to other planned future experiments like the DUNE near detector, PIONEER, Hyper-K for $m_N < m_K$, and SHADOWS for $m_N > m_K$.
It is however not clear to us, however, how this can be achieved, since Mount Thor presents the largest vertical cliff on Earth.

Regarding the SPYGLASS satellite we have found that its sensitivity is likely to be very significant compared to existing limits and future projections, and it strongly depends on the lower energy threshold of the signal of appearing charged particles. 
Already with a lower energy threshold of 100~GeV this experiment would cover a large, unprobed portion of HNL parameter space, while a threshold as low as 10~GeV (similar to the PAMELA threshold) would constitute the first stress-test of the seesaw mechanism. Concretely, SPYGLASS 10 would detect HNL with active-sterile mixing parameters $|U_\alpha|^2$ as small as ${\cal O}(10^{-11})$ for $m_N \sim$ 300 MeV, while also giving the leading constraints in the 0.1 - 2 GeV mass range for electron coupling dominance, and in the whole 10 MeV - 2 GeV range examined in this article for the muon dominance scenario.

We remark that our results are obtained under the assumption that possible backgrounds are completely reducible and will be well under experimental control. The `background-free hypothesis' is supported for MtThor by the analysis of sub-horizon observations at SHALON, and for SPYGLASS by the geometry of its setup and the background rejection capabilities of the PAMELA satellite at energies above 10 GeV.

We conclude that the idea of cosmic-ray beam dump experiments and the promising sensitivity projections warrant the exploration of dedicated experiments searching for atmospherically produced other light neutral LLPs, among others: light scalars, pseudoscalars (including Axion-Like Particles) and new vector bosons (dark photons, light Z' mediators) among other interesting scenarios. We leave the study of these and other LLP realizations for future work.

\section*{Acknowledgements}
We would like to thank Anatoli Fedynitch for technical support with the \mceq software.  BP and JZ are supported by the {\it Generalitat Valenciana} (Spain) through the {\it plan GenT} program (CIDEGENT/2019/068), by the Spanish Government (Agencia Estatal de
Investigación) and ERDF funds from European Commission (MCIN/AEI/10.13039/501100011033, Grant No. PID2020-114473GB-I00).

\bibliographystyle{JHEP} 
\bibliography{cosmic_beam_dumps}

\end{document}